# A new experimental technique for investigation of plasma generated with plasmotrons in electrophysical installations

V.Yu. Khomich, I.I. Kumkova, Yu.A. Zheleznov

*Scientific Instrumentation Center of Institute of Problems of Electrophysics of the Russian Academy of Sciences*

**Abstract**

A new experimental technique for investigating characteristics of plasma generated with plasmotrons in electrophysical installations was proposed. The technique involves a simultaneous registration of both radiation spectra and images of plasma and consequent analysis of the registered data [1].

The experimental technique proposed permits to acquire much more information about plasma characteristics compared to common techniques of separate registration of spectra and images and reveal possible correlation between spectral parameters and image features of plasma [2].

The technique proposed was experimentally realized using complex of apparatus fulfilling simultaneous registration of both radiation spectra and images of plasma generated in electrophysical installations designed for destruction of hazardous and harmful substances [3].

Analysis of the data acquired using the proposed technique permits to find both optimum operating conditions of existing plasmochemical reactors and the optimum number and spatial arrangement of plasmotrons in reactors being designed. Optimization here means attaining maximum effectiveness of destruction of the substances, increasing destruction speed, reducing consumption of electric power and consumable chemicals, and a whole number of the other positive results.

It is also possible to use output signals of the registering apparatus to control plasmotrons' operating conditions in real time i.e. to build plasmochemical reactor with feedback.

## 1. Introduction

In all over the world a fast growth of various wastes (industrial, medical, biological etc.) that have to be destructed or recycled takes place nowadays. Every industrially developed country puts out considerable efforts to find the most optimal technologies to solve this global and very complicated problem.

Methods of thermal waste treatment have been widely used in the past decades [4, 5]. The thermal treatment methods have a whole series of advantages over other methods of waste destruction and decontamination, however they also have serious drawbacks such as insufficient degree of waste destruction and large volume of hazardous substances ejected in Earth's atmosphere. This causes necessity of building purification plants that are bulky, complicated and expensive in both construction and maintenance.

The most promising way of treatment and decontamination of hazardous and harmful wastes today is using plasmochemical technologies that have attracted much attention in all industrially developed countries lately [6, 7].

Practically all electrophysical installations designed for treatment of various wastes use high current arc discharges and electric arc generators of low temperature plasma – plasmotrons [8, 9].

Plasmotron allows to generates dense plasma with temperature in 5000÷10000 K range, stabilize it in space and practically use it any gas environment. Plasmotron power vary in a very wide range – form dozens of watt to dozens of megawatt. In palsmotron internal energy of electric arc transforms in internal energy of gas surrounding the arc.

Main tasks in the area of designing and building electrophysical installations using plasmochemical technologies are increasing both effectiveness of waste destruction and total efficiency, reducing consumption of electric power and consumables, increasing service life and capacity, decreasing maintenance and repair costs etc. To succeed in the above it is necessary first of all to investigate characteristics of plasma generated with plasmotrons in plasmochemical reactors.

## 2. A new technique for investigation of plasma

Many techniques of plasma diagnostics have been developed to date. These techniques fall into two major types – passive, or non-contact, and active.

Active techniques use probes and sensors introduced into plasma, laser radiation, probing particle beams.

Passive diagnostics techniques use processes and phenomena that take place in plasma itself and can be registered at its periphery: electromagnetic radiation in wide wavelength range (X-ray, ultraviolet, visible, infrared, microwave), corpuscular emission, electric and magnetic fields existing around the plasma region.

Passive techniques have an important advantage of not disturbing the plasma in contrast to active techniques.

In spite of a variety of existing techniques they are not completely satisfied for diagnostics of plasma generated with plasmotrons in electrophysical installations because they do not allow receiving sufficient information about processes occurring in the plasma through complexity and diversity of processes taking place in the plasma. There is a whole series of chemical reactions going in the plasma. The situation is complicated by the fact that chemical and physical processes in the plasma are not independent.

The authors developed a new approach to examination of plasma properties. An idea was advanced to fulfill plasma diagnostics by simultaneous use of two optical techniques – registration of both radiation spectra and images of plasma. Optical techniques were selected because they are the most informative among passive ones used for research of low temperature plasma. By using optical techniques it is possible to conduct measurements of practically all plasma parameters in full scale.

Such the synchronous diagnostics will permit to acquire much more information about plasma characteristics compared to techniques of separate registration of spectra and images and may reveal possible correlation between spectral parameters and image features. Moreover simultaneous registration of both spectra and images requires less time compared to registration of spectra and images in series.

## 3. Selection of spectrum registering apparatus

Combination of imaging spectrograph and scientific grade CCD camera was selected to register optical spectra of plasma's radiation.

Scientific grade CCD cameras in contract to consumer ones feature very low readout noise and dark current signal. Photons of radiation falling onto photosensitive chip of CCD camera produce photoelectrons that are confined to their respective elements of the chip. Thus if a spectrum (or any light pattern) is projected onto the chip, a corresponding charge pattern will be produced. This charge pattern then is transferred off the chip into the shift register, amplified and passed via control/readout cable to analog-digital converter located on interface board, which is installed in computer. The spectrum is then recorded in file for further analysis and archiving.

There are two most common types of CCD cameras used nowadays: Si (silicon) based and InGaAs (indium gallium arsenide) based. They differ in spectral sensitivity range – Si CCDs cover 400-1100 or 180-1100 nm range, and InGaAs ones cover 800-1700 nm range.

Range of CCD's spectral sensitivity depends on its design. In front-illuminated CCDs photons fall on the chip through electrode structure. Absorption in the electrode structure prevents detection of wavelengths below 400 nm. Ultraviolet coating (phosphor) can be put on front-illuminated device to give ultraviolet response but it has lower quantum efficiency and can be bleached by high fluxes of ultraviolet radiation.

An alternative to ultraviolet coating is to thin the silicon chip down, from the back, to 10-15 μm (similar thickness to the sensitive region) and then illuminate it from the back. Such the design of CCD is called back-illuminated. Spectral sensitivity range of back-illuminated devices is extended down to 180 nm. The main drawback of such the design is that it is expensive and device can suffer from fringing.

CCD can operate in different modes achieving different spectra registration rate. In full resolution mode the maxim rate is few spectra per a second only. To increase the rate full vertical binning mode is used. In this mode all the rows are transferred into the shift register before the charge is transferred from the shift register into the amplifier. Binning provides several advantages such as increased readout speed and better noise performance. In full vertical binning mode about few hundreds spectra per a second can be registered.

To increase spectra registration speed it is necessary to use combination of CCD and imaging spectrograph. The term "imaging" is used in spectroscopy instrumentation to describe the point-to-point replication of the input slit at the output plane of the spectrograph. Standard non-imaging spectrographs will only produce a one-dimensional spectrum at their focal plane, i.e. there is no resolution or imaging ability perpendicular to the dispersion. Imaging spectrographs however have specially designed toric mirrors that allow the imaging of the entrance slit on to the focal plane.

If one inputs light into the imaging spectrograph with the 100 μm diameter optical fiber then about 10 rows only of the CCD chip will be exposed to light. The rows can be shifted with speed of up to 1 μsec per a row. So speeds of up to $10^5$ spectra per a second can be attained. Another advantage of imaging spectrograph is that several spectra can be collected from different space points simultaneously. Inputting light into spectrograph with multi-track fiber bundle does it.

To further increase spectra registration rate it is necessary to use combination of so called fast kinetics CCD and imaging spectrograph. The fast kinetics CCD has photosensitive chip with mask applied on the chip. The mask is applied so that only one exposed row is left close to the top of the CCD and 128 exposed rows at the bottom of the CCD. Rows are shifted down until all the non-illuminated area below the exposed row is filled. After that a readout process is occurred. For this type of CCD speed of up to $10^6$ spectra per a second is achievable given 1 μsec per a row vertical shift speed.

Two CCD cameras were selected for our experiments. The first one is back-illuminated CCD sensitive in 180-1100 nm range. Its format is 1024x256 pixels of 26x26 μm$^2$ size. The second one is fast kinetics CCD. It has the same characteristics and design as the first CCD except of the mask. The metal mask is applied so that 1 exposed row is left close to the top of the CCD and 128 exposed rows at the bottom of the CCD. The open lower half allows the CCD to be used for standard measurements as well. An imaging spectrograph was also selected for the experiments.

## 4. Selection of image registering apparatus

Parameters of plasma generated with plasmotrons in plasmochemical reactors change in time very fast. It necessitates use of high-speed apparatus to register plasma images. Contemporary apparatus for high-speed image registration fall into two major categories: electronic imaging cameras and mechanical cameras.

The distinctive element of electronic imaging cameras is an image converter tube that converts photons to an array of electrons analogous to the image. This electron image is electrostatically focused, deflected and shuttered to produce a record. The non-mechanical manipulation of the image allows for extremely fast recording speeds and provides great flexibility in the timing and duration of exposures. A phosphor screen in the image converter tube reproduces the optical image by converting electrons back to photons that are recorded on film or to charge-coupled device.

Mechanical cameras use moving parts and optics to form an image on the film. The central part of most mechanical cameras is a mirror that rotates at a very high speed, reflecting light from the subject to the film track. Because of the accurate timing requirement associated with high speed photography, the timing of the event and exposure is performed by support equipment. Conventional shutters are too slow to time the actual exposure, but are useful for preventing film fogging and exposing static images for reference. A serious drawback of mechanical cameras is a mechanical wear.

In both electronic imaging and mechanical cameras a photographic film or CCD can be used as a light registering media. Film's serious disadvantage is that it is a nonrecoverable registering media. Moreover it is necessary to fulfill time-consuming procedures of developing the film and digitizing the images for further computer analysis.

Charge-couple devices are free from film's disadvantages. Drawback of CCD systems is that at very high registration speeds of about a million frames per a second a separate CCD must be used to register one frame in a series. Therefore the total number of frames in a series is limited by both technical reasons and price.

It is also possible to register images directly with CCD camera. In this case light falls on photosensitive chip through a lens that projects image of an object on the chip. To use full resolution capability of the CCD it is necessary to transfer a row at a time into the shift register and then to readout the shift register. This way is cheaper compared to using either image converter tube or rotating mirror equipment. Unavoidable sacrifice here is decrease of frame rate and increase of exposure time.

To achieve a very short exposure time intensified charge-couple device can be used. It is a combination of an image intensifier and CCD coupled together with fiber optic or lens. A photon strikes photocathode of the image intensifier producing a photoelectron that is drawn towards a micro channel plate. The photoelectron cascades down the channel producing secondary electrons and exiting as a cloud of electrons. Resultant amplifications can be up to $10^4$. If the voltage of the photocathode is made positive relative to the input of micro channel plate, then no photoelectrons will be emitted and the image intensifier will effectively be off. Suitable gating electronics can switch between "on" and "off" levels in < 1 nsec, resulting in optical gate "on" periods of as low as 2 nsec to infinity.

Disadvantage of intensified CCD is a low frame rate because a CCD is used as registering device. Another disadvantage is a reduced range of spectral sensitivity that is defined by photocathode material.

For our experiment high speed camera with direct registration of images on CCD chip was selected. The camera is sensitive in 400-1000 nm range. Chip format is 652x496 pixels. The chip is splitted in four quadrants so that every quadrant is read out in its shift register. Such the design increases frame rate by four. Recording speed is variable with maximum of about several thousand frames per a second at lowered resolution of 20x40 pixels. The images are recorded directly into the camera's memory and stored on computer's hard disk. The camera is equipped with electronically controlled zoom lens.

Intensified CCD was also selected for plasma image registration. Its photocathode is sensitive in 180-850 nm range, optical gate < 5 nsec.

## 5. Scheme of the experiment on simultaneous registration of plasma's optical radiation spectra and images

Scheme of experiment on synchronous registration of plasma spectra and images is shown in figure 1.

Optical radiation of plasma is withdrawn out of plasmochemical reactor via diagnostics window. Then radiation is fallen onto beamsplitter where it is divided in two beams. The first beam is directed in spectra registration channel, the second one is directed in image registration channel.

The first beam gets into the spectrograph through various removable input modules including fixed and variable width slits, single fibers, multiple-track fiber bundles. Variety of input modules provides flexibility of the experiment and ensures operation in various regimes.

Then light passes through one of filters installed in a motorized filter wheel. Various types of filters can be used here.

Light is diffracted from the spectrograph's grating in several diffraction orders. Usually only the first order, positive or negative, is desired. The other wavelengths in higher orders may need to be blocked. It is done by using order sorting filters.

It may also be necessary to view several or even one specific wavelength only. Broadband or narrow bandpass interference filters can be used for that.

The spectrograph is equipped with multiple-grating turret that holds several diffraction gratings simultaneously. It ensures flexibility of the experiment by allowing registration of different parts of spectrum with different spectral resolution.

All the functions of imaging spectrograph, CCD camera and high-speed camera (selection of grating, filter, spectral range, exposure time, binning scheme, the number of spectra and images in a series etc.) are software controlled.

Synchronous operation of spectral and image channels is provided with trigger pulse that is sent from the CCD camera's interface board. The trigger pulse is transmitted from interface board directly to CCD camera and via input-output unit to high-speed camera's control unit.

Spectrograph's mechanical shutter is also controlled from interface board. The shutter is needed to register background spectrum that is subtracted from signal spectra. The shutter is also used when light gets into imaging spectrograph with multi-track fiber bundle. In this case several spectra are registered simultaneously. If light was to continue to fall on the CCD chip during the charge transfer process, the individual tracks would sample light from more then their intended positions and thus the resultant spectra would appear streaked.

Selection of diffraction grating, rotating of a grating to set desired spectral range, and wavelength calibration are done using IEEE connection.

One more scheme of registering plasma's spectra and images was also proposed. The scheme is depicted in figure 2. Here the second diagnostics window is added. Advantage of this scheme compared to one-window scheme is absence of optical distortions introduced by beamsplitter. Another advantage is absence of beam attenuation that is caused by both light absorption in beamsplitter material and dividing the beam in two parts by beamsplitter.

Drawback of two-window scheme is increase of plasma perturbations introduced by the windows. Moreover adding the second window increases construction and maintenance costs.

It is necessary to do wavelength calibration of spectral channel. Such the calibration is done with spectral calibration light sources based on Hg(Ar) and Ne lamps and radiometric power supply.

In the case of necessity of measuring absolute intensities of spectral lines it is also necessary to do irradiance calibration of spectral channel. This procedure is more time consuming and complicated than wavelength calibration. In addition any change in optical scheme e.g. introducing filter in optical path requires making re-calibration. Irradiance calibration is done using calibrated irradiance sources. QTH lamp based source provides calibration in 250-2500 nm range, and deuterium lamp based source in 200-400 nm range.

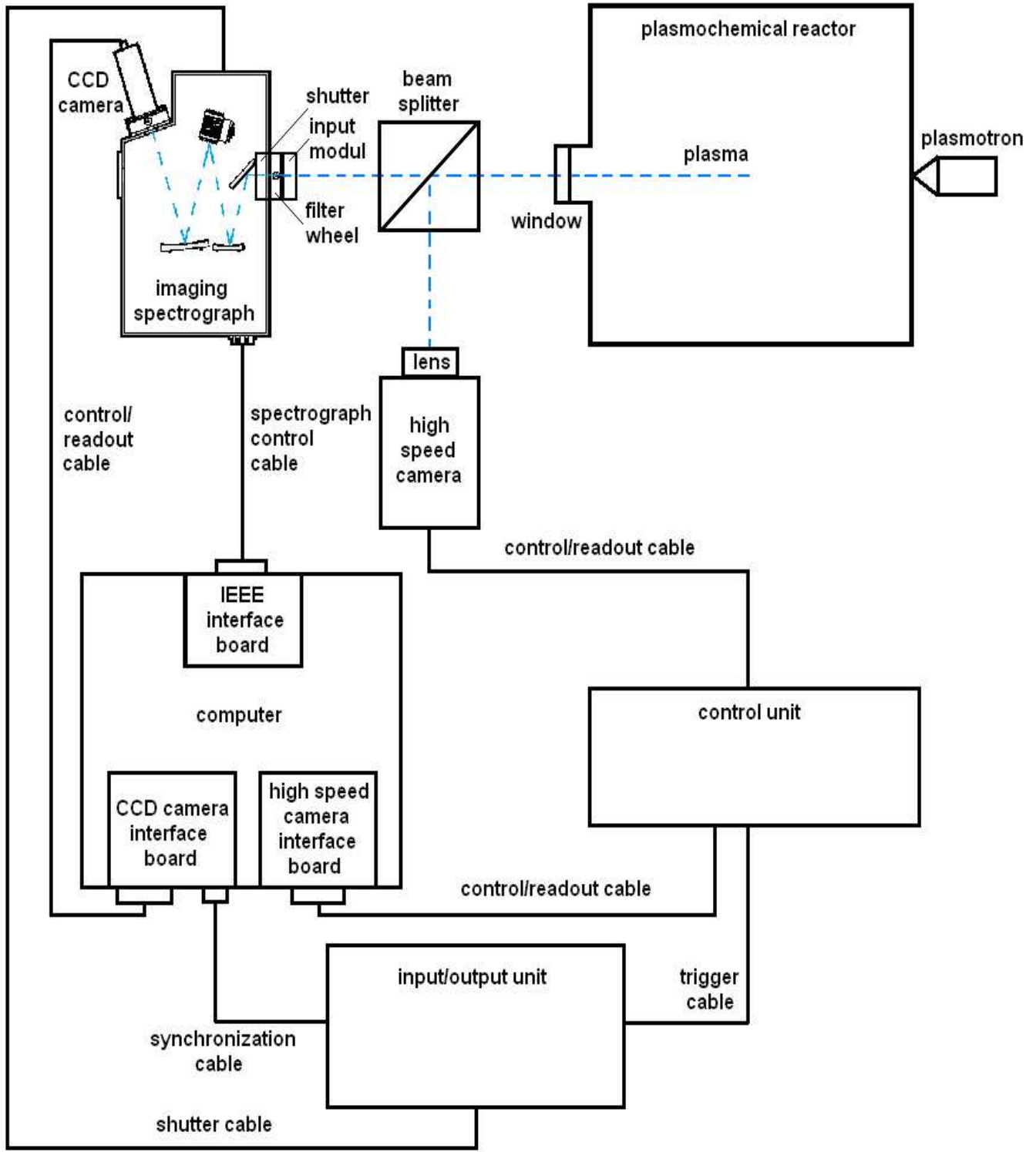

**Figure 1.** Scheme of experiment on synchronous registration of plasma spectra and images with one diagnostics window.

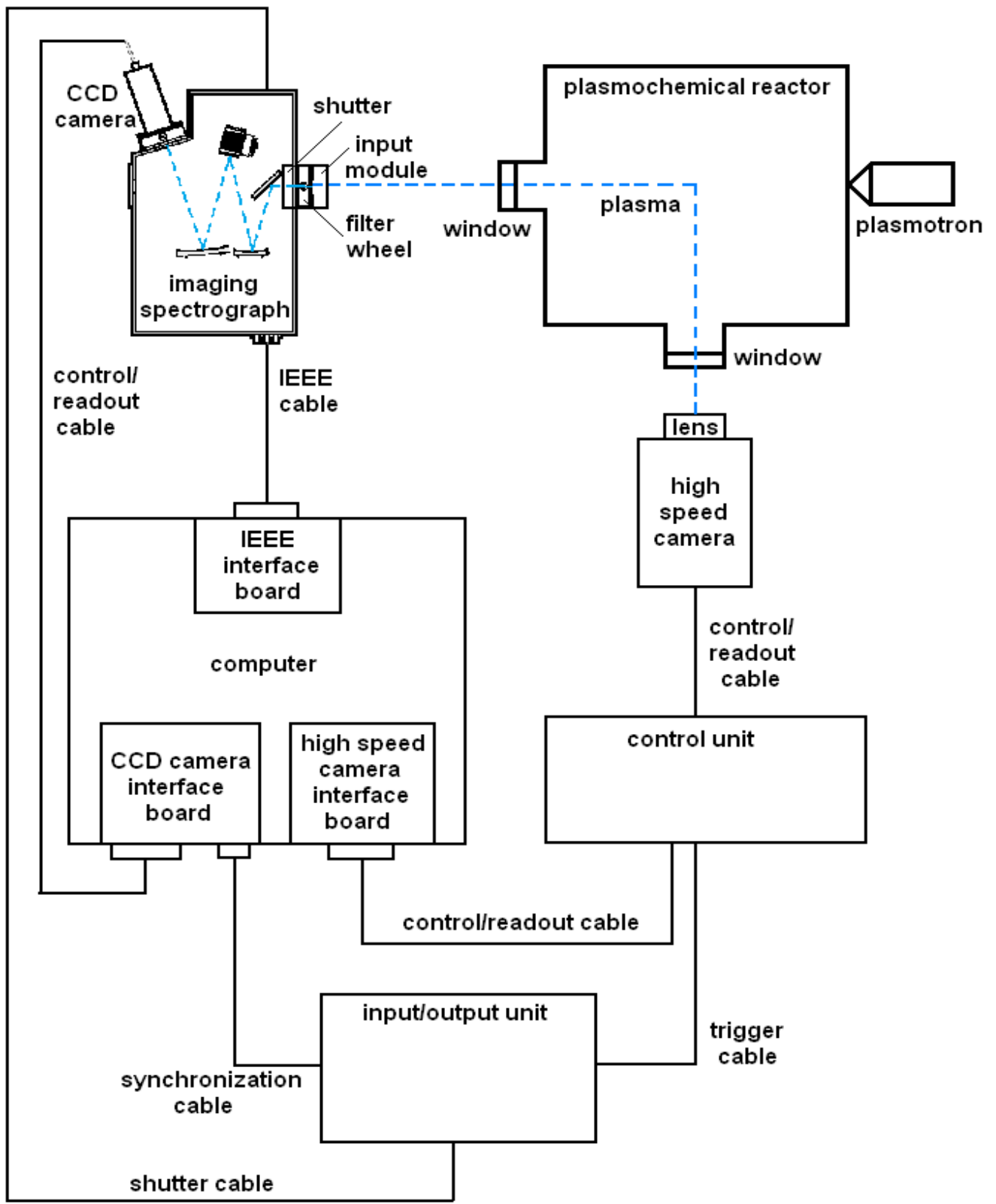

**Figure 2. Scheme of experiment on synchronous registration of plasma spectra and images with two diagnostics windows.**

# REFERENCES

[1]   V.Yu. Khomich, Yu. A. Zheleznov. A new technique of investigating plasma characteristics. // Preprint of Scientific Technology Center of Unique Instrumentation of RAS. Moscow. 2001.

[2]   V.Yu. Khomich, Yu. A. Zheleznov. Experimental techniques and complex of apparatus for investigation of plasma generated with plasmotrons in electrophysical installations designed for environment protection. // Preprint of Scientific Instrumentation Center of Institute of Problems of Electrophysics of RAS. Moscow. 2002.

[3]   Ph.G. Rutberg. Plasma Pyrolysis of Toxic Waste. // Plasma Physics and Controlled Fusion 45 (2003) 957-969.

[4]   Eds R.E.Hester and R.M.Harrison. Waste Incineration and the Environment. // Manchester. 1994.

[5]   Recommendations for the Disposal of Chemical Agents and Munitions. // National Research Council. 1994.

[6]   G.W. Carter, A.V.Tsangaris. Plasma gasification of biomedical waste. // Proc. Intern. Symp. on Environment Technol. Plasma Systems and Applications, 8-11 Oct. 1995, Atlanta, Georgia, USA. v.1.

[7]   Recommendations for the Disposal of chemical Agents and Munitions. National Research Council. 1994.

[8]   A.S. Koroteev, V.M. Mironov, Yu.S. Svirchuk. Plasmotrons: designs, characteristics, computations. // Moscow. Mashinostrojenie. 1993.

[9]   I.A. Glebov, Ph.G.Rutberg. High-power plasma generators. // Moscow. Energoatomizdat. 1985.